\documentclass{article}
\usepackage{spconf,amsmath,graphicx}

\usepackage{enumitem}
\setlist{nosep, leftmargin=14pt}

\usepackage{mwe} 

\usepackage{multirow} 
\usepackage{gensymb} 
\usepackage{amsfonts} 
\usepackage{amssymb} 
\usepackage{float} 
\usepackage{makecell} 
\usepackage[dvipsnames]{xcolor} 

\usepackage{amsmath,amsfonts,bm}

\def\eqref#1{equation~\ref{#1}}

\def\1{\bm{1}}

\DeclareMathAlphabet{\mathsfit}{\encodingdefault}{\sfdefault}{m}{sl}
\SetMathAlphabet{\mathsfit}{bold}{\encodingdefault}{\sfdefault}{bx}{n}

\newif\ifshowcomments
\showcommentstrue 

\usepackage[dvipsnames]{xcolor}
\newcommand{\shortname}{Epi-NAF\xspace}
\ifshowcomments
    \newcommand{\comment}[1]{\textcolor{olive}{{\em #1}}}
    \newenvironment{multilinecomment}[1]{\begingroup\color{olive}#1}{\endgroup}
    \newcommand{\TK}[1]{\textcolor{orange}{{\em {\bf TK:} #1}}}

    \newcommand{\oneline}[1]{{\textcolor{Emerald}{\textbf{One-Liner:} {#1}}}}
\else
    \newcommand{\comment}[1]{}
    
    \newcommand{\TK}[1]{}
    
    \newcommand{\oneline}[1]{}
\fi

\title{EPI-NAF: Enhancing Neural Attenuation Fields for Limited-Angle CT with Epipolar Consistency Conditions}
 \name{Daniel Gilo$^{1}$\qquad Tzofi Klinghoffer$^{2}$ \qquad Or Litany$^{1, 3}$}
 \address{$^{1}$Technion – Israel Institute of Technology \\
     $^{2}$Massachusetts Institute of Technology \\
     $^{3}$NVIDIA}
\begin{document}
%
\maketitle
\begin{abstract}

Neural field methods, initially successful in the inverse rendering domain, have recently been extended to CT reconstruction, marking a paradigm shift from traditional techniques. While these approaches deliver state-of-the-art results in sparse-view CT reconstruction, they struggle in limited-angle settings, where input projections are captured over a restricted angle range. We present a novel loss term based on consistency conditions between corresponding epipolar lines in X-ray projection images, aimed at regularizing neural attenuation field optimization. By enforcing these consistency conditions, our approach, \shortname, propagates supervision from input views within the limited-angle range to predicted projections over the full cone-beam CT range. This loss results in both qualitative and quantitative improvements in reconstruction compared to baseline methods.
\end{abstract}
\section{Introduction}
\label{sec:intro}

In 3D limited-angle computed tomography (LACT), the goal is to reconstruct a radiodensity voxel grid from X-ray projection images acquired over a restricted range of angles. This inverse problem is ill-posed and presents significant challenges in various applications, such as electron microscopy \cite{quinto2008local}, breast tomosynthesis \cite{zhang2006comparative}, C-arm CT \cite{schafer2012limited}, and non-destructive testing \cite{krimmel2005discrete}.

Since the 1970s, several paradigms for CT reconstruction have been developed. Traditional   analytical 
reconstruction methods, such as the Feldkamp-Davis-Kress (FDK) algorithm \cite{feldkamp1984practical}, often introduce significant artifacts into reconstructed images in limited-angle scenarios. Iterative reconstruction techniques, such as ART \cite{gordon1970algebraic} and SART \cite{andersen1984simultaneous}, along with more modern variations incorporating image priors \cite{sidky2006accurate, niu2014sparse, huang2013iterative}, provide improvements by reducing artifacts and enhancing image fidelity. In recent years, deep learning-based methods have emerged as powerful tools for addressing the challenges of LACT, achieving notable advancements in reconstruction quality \cite{liu2023dolce, chen2022sam}. However, these methods often operate on 2D slices due to the scarcity of 3D CT data.

Inspired by the success of Neural Radiance Fields (NeRF) \cite{mildenhall2021nerf} in inverse rendering, a new family of approaches represents the attenuation coefficient volume as a \emph{neural field}: a continuous function of 3D coordinates, parameterized by neural networks \cite{fang2022snaf, cai2024structure, zha2022naf, zang2021intratomo, ruckert2022neat}. Neural field-based approaches optimize the scene representation to match a given set of input views by leveraging the differentiability of the rendering process. The representation is typically optimized seperately for every scene. These approaches produce state-of-the-art results in the sparse-view CT setting. However, in the limited-angle setting, certain regions of the reconstructed volume may remain significantly underconstrained, often resulting in blurry or low-quality reconstructions.

To address this limitation, we propose a novel regularization term for neural field optimization in the LACT setting, based on epipolar consistency conditions in X-ray imaging \cite{aichert2015epipolar}. These conditions, derived from the Grangeat theorem \cite{grangeat1991mathematical}, enforce consistency between the derivatives of line integrals along corresponding epipolar lines, across arbitrary epipolar geometries.
Our key insight is that these terms can be leveraged to constrain neural field optimization with respect to 
projections
both within and outside the limited-angle region. By enforcing these conditions across the entire 180\degree\ range\footnote{For consistency with previous works we assume semicircular acquisition trajectory, but the extension to 360\degree\ is trivial.}, our proposed approach, \shortname, effectively \emph{propagates supervision} from input views within the limited-angle range to unseen angles. This serves as a powerful complementary loss term alongside the standard neural field 
loss.

In the following sections, we describe our method and demonstrate that it outperforms both traditional and neural field-based methods, both quantitatively and qualitatively, across a range of CT scans and limited-angle configurations.

\begin{figure}[h!]
    \centering
    \includegraphics[width=1.0\linewidth]{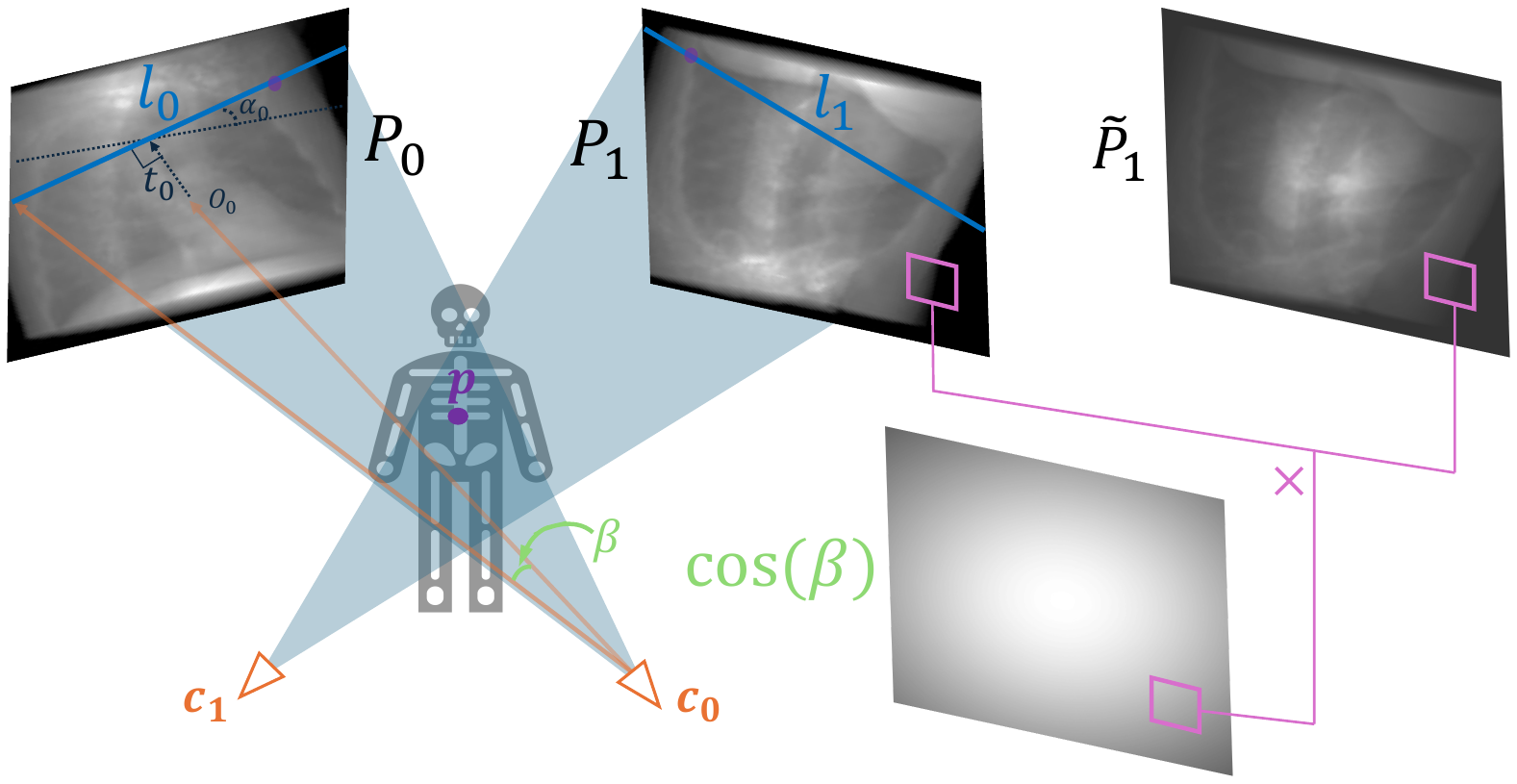}
    \caption{\textbf{Epipolar consistency in X-ray imaging.} $P_0$ and $P_1$ are two X-ray projection images, where \textcolor{orange}{$\boldsymbol{c}_0$} and \textcolor{orange}{$\boldsymbol{c}_1$} are the corresponding X-ray source locations. We consider the cosine-weighted projection images, $\Tilde{P_0}$ and $\Tilde{P_1}$, obtained by \textcolor{Thistle}{multiplying} the intensity value of each pixel by \textcolor{LimeGreen}{$cos(\beta)$}, where \textcolor{LimeGreen}{$\beta$} is the angle between the pixel, the source location and the image origin $O$. The source locations, along with a point in the body \textcolor{violet}{$p$} define an epipolar plane, which intersects the projection images at \textcolor{RoyalBlue}{$l_0$, $l_1$}. The two corresponding epipolar lines are defined by the angle $\alpha$ and the distance to the origin $t$.}
    \label{fig:epipolar}
\end{figure}

\section{Background}
\label{sec:background}

\subsection{Neural attenuation fields}
Neural Attenuation Fields (NAF) \cite{zha2022naf} is a method inspired by NeRF for reconstructing the attenuation coefficient field of a scene from a limited number of 2D X-ray projections in a cone-beam computed tomography (CBCT) setting. In NAF, a multilayer perceptron (MLP) $\phi$ is optimized to model a function that maps a 3D spatial coordinate to its corresponding attenuation coefficient, $\mu$. 

For an X-ray source location $\boldsymbol{c}$ along a semicircular source trajectory, and a direction $\boldsymbol{d}$, the predicted intensity value of a ray $\boldsymbol{r}(s) = \boldsymbol{c} + s\boldsymbol{d} \in \mathbf{R}^3$ is computed using the discretized form of the Beer-Lambert law:

\begin{equation} \hat{I}(\boldsymbol{r}) = I_0 \exp{\left(-\sum_{i=1}^{N} \mu_i \delta_i\right)}, \label{eq:beer}\end{equation}

where $I_0$ is the initial intensity, $\mu_i = \phi(\boldsymbol{r}(s_i))$ represents the predicted attenuation coefficient at the $i^{th}$ sampled point along the ray, and $\delta_i = \lVert \boldsymbol{r}(s_{i+1}) -\boldsymbol{r}(s_{i}) \rVert$ denotes the distance between consecutive sampled points.

The network parameters are optimized by minimizing the $L_2$ difference between the predicted projections and the given input projections over the training ray batch $\Omega_{recon}$:

\begin{equation} \mathcal{L}_{\text{recon}} = \frac{1}{|\Omega_{recon}|}\sum_{\boldsymbol{r}\in \Omega_{recon}} \lVert \hat{I}(\boldsymbol{r}) - I(\boldsymbol{r}) \rVert_2^2 \end{equation}

After optimization, the model can be queried with the coordinates of cells in a voxel grid of arbitrary resolution to generate the reconstructed CT scan.

Although NAF demonstrates strong performance in the \emph{sparse-view} CT setting—producing high-quality reconstructions from several dozen input projections spanning a 180\degree\ range—it performs poorly in the limited-angle setting. In this case, the reconstructed $\mu$ field is insufficiently regularized in certain regions, leading to degraded reconstruction quality.

\begin{figure*}
    \centering
    \includegraphics[width=0.9\linewidth]{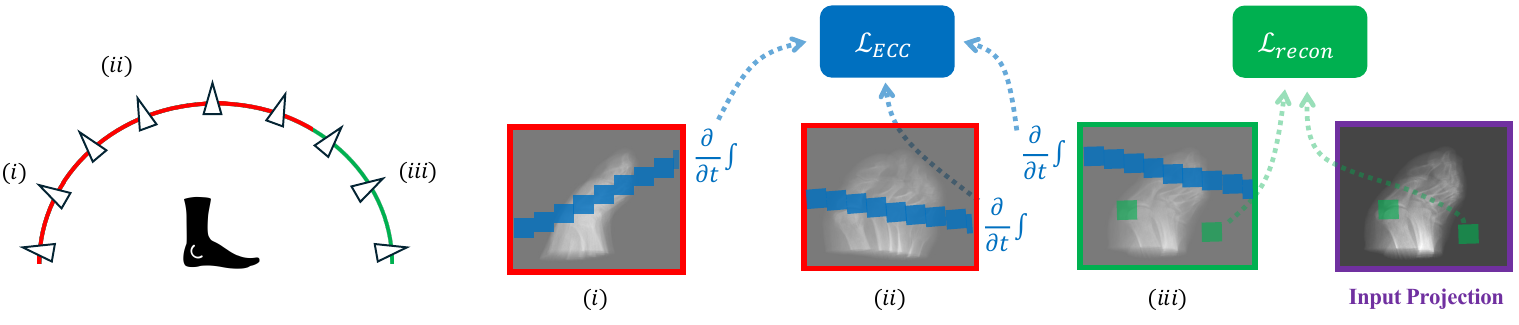}
    \caption{\textbf{Epi-NAF Method Overview.} On the left part of the figure, we mark the limited-angle region of the provided projections in \textcolor{Green}{green}, and the unseen projections region in \textcolor{red}{red}. 
    \shortname comprises two loss terms: (1) $\mathcal{L}_{\text{recon}}$, based on the $L_2$ difference between the predicted and ground-truth intensity values (\textcolor{Green}{green pixels}), and (2) our novel $\mathcal{L}_{\text{ECC}}$, which enforces consistency in the derivatives in the $t$ direction of line integrals along corresponding epipolar lines (\textcolor{RoyalBlue}{blue pixels}). Crucially, projection $(iii)$ receives direct supervision from the \textcolor{Plum}{input}, which is propagated to projections $(i)$ and $(ii)$ via the ECC loss. This propagation of supervision from limited input angles to unseen projections enhances the overall reconstruction quality.
    }
      \label{fig:method}
\end{figure*}

\subsection{Epipolar consistency conditions in X-ray projections}\label{sec:ecc}

Our approach leverages epipolar consistency conditions (ECC) in X-ray imaging, as introduced by Aichert et al. \cite{aichert2015epipolar}. The authors utilize the Grangeat theorem \cite{grangeat1991mathematical} to establish consistency conditions between corresponding epipolar lines in two X-ray images of the same static object that is fully visible in both views.

We consider the following epipolar setting: Given two projection images, $P_0$ and $P_1$, from X-ray source locations $\boldsymbol{c}_0$ and $\boldsymbol{c}_1$, we consider their cosine-weighted versions, $\Tilde{P_0}$ and $\Tilde{P_1}$, where each pixel is weighted by the cosine of the angle between the pixel, the source, and the image origin\footnote{In \cite{aichert2015epipolar} the weighting is ommited as small angles, and therefore uniform weights, are assumed.}. Let $l_0 = \Tilde{P_0}(\alpha_0, t_0)$ and $l_1 = \Tilde{P_1}(\alpha_1, t_1)$ represent two corresponding epipolar lines on these projection images, where $\alpha_i$ denotes the angle between the line $l_i$ and the horizontal image axis, and $t_i$ represents the signed distance to the image origin. 
The epipolar geometry is presented in Figure \ref{fig:epipolar}.

The ECC in flat-detector CBCT state:

\begin{equation}\label{eq:ecc} \frac{\partial}{\partial t} \mathcal{R}_{\Tilde{P_0}}(\alpha_0, t_0) = \frac{\partial}{\partial t} \mathcal{R}_{\Tilde{P_1}}(\alpha_1, t_1), \end{equation}

where $\mathcal{R}_P(\alpha,t)$ is the 2D Radon transform of the projection image $P$.

In simpler terms, the ECC require that the derivatives of the line integrals along corresponding epipolar lines, taken in directions orthogonal to those lines, must be equal.

While these conditions were originally designed for correcting estimated projection geometry, we repurpose them to regularize NAF optimization.

\section{Method}
\label{sec:method}

We propose a novel loss term based on the ECC, to complement $\mathcal{L}_{\text{recon}}$ and regularize the training of the NAF network $\phi$, particularly for regions that are insufficiently constrained by the input projections.

Our key insight is that, in contrast to $\mathcal{L}_{\text{recon}}$,  which can only be applied to predicted projections where there are ground-truth input views for comparison, our proposed term is applicable across the entire 180\degree\ range.

To compute this loss, for a pair of projection angles and their corresponding source locations $\boldsymbol{c}_0, \boldsymbol{c}_1$, we consider the plane they define along with a point $p$ within the projected body. This epipolar plane intersects the predicted projection images \( P_0 \) and \( P_1 \) at the corresponding epipolar lines \( l_0 = P_0(\alpha_0, t_0) \) and \( l_1 = P_1(\alpha_1, t_1) \), as described in Sec.~\ref{sec:ecc} and visualized in Figure \ref{fig:epipolar}.

To evaluate epipolar consistency as described in Eq.\ref{eq:ecc}, each epipolar line $l_i$ is sampled at $N_s$ points, spaced by a distance $\delta_i$, denoted as $\{\hat{p}_{i,j}\}_{j=1}^{N_s}$. 
We define $\Delta_{i, j} = \frac{\partial}{\partial t} \Tilde{P_i} (\hat{p}_{i,j})$, the derivative of the cosine-weighted predicted intensity image, in the direction orthogonal to the line, evaluated at the $j^{th}$ sampling point along the epipolar line $l_i$.

The epipolar consistency loss is defined as:

\begin{equation}
    \mathcal{L}_{\text{ECC}} = \Vert 
     \sum_{j=1}^{N_s}\Delta_{0,j}\delta_0 -  \Delta_{1,j}\delta_1
    \Vert_2^2,
\end{equation}

which corresponds to the squared difference between the derivatives of the approximated integrals along $l_0$ and $l_1$. This loss term thus enforces adherence to the ECC.

In practice, the derivatives in $\Delta_{i, j}$ are approximated using central difference, therefore the predicted projection images are actually sampled along the lines $l_i^+ = P_i(\alpha_i, t_i +\epsilon)$ and $l_i^- = P_i(\alpha_i, t_i -\epsilon)$.
The projection angles and the point 
$p$, which define the epipolar geometry, are randomly sampled at each training iteration.

The total loss for optimizing the network is a weighted combination of both the reconstruction loss and the epipolar consistency loss:
\begin{equation}
    \mathcal{L} = \mathcal{L}_{\text{Recon}} + \lambda \mathcal{L}_{\text{ECC}},
\end{equation}

where $\lambda$ is a hyperparameter that controls the trade-off between the two loss terms.

Our proposed method is depicted in Figure \ref{fig:method}.

\section{Empirical Evaluation}
\label{sec:evaluation}

\bgroup
\def\arraystretch{1.2}%
\begin{table}[t!]
\setlength{\tabcolsep}{1.5pt}
    \centering
    \small
    \begin{tabular}{cccccc}
         Scan & Method & 45\degree &  60\degree & 90\degree &
         120\degree \\
         \hline 
         \multirow{4}{*}{Abdomen} & {FDK}  & 11.32/.278 & 14.52/.301 & 16.35/.333 & 18.69/.363
         \\
         & SART  & 21.96/.729 & 22.92/.754 & 24.33/.795 & 26.47/.846 
         \\
         & ASD-POCS  &  22.09/.744  &  23.08/.77& 24.51/.813 & 26.75/.867
         \\ 
         &{NAF}   & 23.01/.816 & 25.75/.858 & 28.582/.892 & 30.46/.919\\
         
        & \shortname
         & \textbf{23.88}/\textbf{.823} & \textbf{26.1}/\textbf{.863} & \textbf{28.81}/\textbf{.897} & \textbf{30.51}/\textbf{.921} \\
         \hline
         
         \multirow{4}{*}{Chest} & {FDK} & 11.54/.191 & 12.87/.273 & 14.96/.419 & 17.97/.514
         \\
         & SART & 17.08/.504 & 17.99/.57 & 21.87/.775 & 24.85/.859
         \\
         & ASD-POCS  & 17.2/.508 &  18.11/.573&  22.05/.78&  25.09/.866
         \\ 
         &{NAF} & 19.1/.591 & 20.4/.678 & 24.78/.834& 27.84/.903\\
         
        & \shortname
        & \textbf{19.6}/\textbf{.606} & \textbf{21.57}/\textbf{.724} & \textbf{25.73}/\textbf{.856} & \textbf{28.28}/\textbf{.909} \\
          \hline
          
        \multirow{4}{*}{Foot} & {FDK} & 16.07/.217 & 16.99/.251 & 18.78/.338 & 20.67/.382
         \\
         & SART & 20.9/.604 & 22.39/.725 & 24.61/.873 & 26.05/.902 
         \\
         & ASD-POCS  & 20.93/.603 & 22.42/.723 & 24.58/.867 & 25.99/.896
         \\ 
        &{NAF} & 24.7/.835 & 25.42/.862 & 26.48/.896 & 28.43/.914\\
        
        & \shortname
         & \textbf{24.9}/\textbf{.849} & \textbf{25.44}/\textbf{.869} & \textbf{26.49}/\textbf{.898} &  \textbf{28.48}/\textbf{.916} \\
         
         \hline
         \multirow{4}{*}{Jaw} & {FDK}  & 19.4/.425 & 20.84/.477 & 22.98/.553 & 25.08/.645
         \\
         & SART & 25.53/.759 & 25.83/.776 & 27.42/.822 & 29.76/.88 
         \\
         & ASD-POCS  & 25.4/.761 & 25.91/.777 & 27.55/.827 & 29.94/.883
         \\ 
         &{NAF}  & 26.29/.765 & 26.89/.787 & 28.78/\textbf{.847}& 30.86/.891\\
         
        & \shortname
         & \textbf{26.36}/\textbf{.772} & \textbf{26.91}/\textbf{.791} & \textbf{28.9}/\textbf{.847} & \textbf{31.21}/\textbf{.9} \\
         \hline
    \end{tabular}
    \caption{PSNR/SSIM compared to ground-truth CT scans, of \shortname and baselines. Best results are in \textbf{bold}.}
    \label{tab:results}
\end{table}
\egroup

\noindent\textbf{Datasets.}
To evaluate the performance of our method, we used the four publicly available CT scans used in NAF~\cite{zha2022naf}, including a chest CT from the LIDC-IDRI dataset \cite{armato2011lung}, and CT scans of a foot, jaw, and abdomen from the Open Scientific Visualization Datasets\footnote{https://klacansky.com/open-scivis-datasets/}. Using the TIGRE toolbox \cite{biguri2016tigre}, we generated 2D X-ray projection images for various limited-angle settings: 45\degree, 60\degree, 90\degree, and 120\degree, with 50 projections per angle range for all datasets.

\vspace{2mm}
\noindent\textbf{Implementation details.} To ensure a fair comparison, we followed the implementation details from the original NAF paper. 
For \shortname 
we set the the loss weight term  $\lambda$ to $10^{-3}$ for angle ranges below 120\degree\ and $10^{-4}$ for 120\degree, as the less ill-posed setting requires less regularization. In every training iteration we calculated $\mathcal{L}_{\text{ECC}}$ for a single pair of randomly-selected epipolar lines. When sampling epipolar lines, the number of sampling points $N_s$ was set to match the number of pixels in the horizontal axis of the detector (ranging from 256 to 1024). We trained for 3000 epochs, with a 200-epoch warm-up period where we only optimized $\mathcal{L}_{\text{Recon}}$. 
Training a NAF with the additional rays required for $\mathcal{L}_{\text{ECC}}$ took between 20 and 180 minutes depending on the scan resolution, on a single NVIDIA GeForce RTX 4090 GPU. This represents a computational cost increase by a factor of 1.5 to 3 compared to the original NAF.

For evaluation, we queried the network to generate a voxel grid with a resolution matching the ground-truth scan (ranging from $128^3$ to $512^3$), representing the reconstructed CT scan. We then compared the reconstruction to the ground-truth CT using standard PSNR and SSIM metrics.

\vspace{2mm}
\noindent\textbf{Baselines.} We evaluated our approach against four baseline methods representing different paradigms for CT reconstruction: the FDK \cite{feldkamp1984practical} method, a widely-used analytical approach adapting filtered back-projection to cone-beam geometry; the iterative SART \cite{andersen1984simultaneous} method; ASD-POCS \cite{sidky2008image}, which incorporates a total variation prior; and the vanilla NAF, representing neural field-based approaches. We used TIGRE's default implementation of both FDK, SART and ASD-POCS.

\subsection{Results and discussion}
As shown in Table \ref{tab:results}, our proposed approach consistently improves the quantitative performance of the vanilla NAF method across all tested scans and limited-angle configurations. Both neural field-based methods outperform traditional techniques in these challenging settings, highlighting the potential of this emerging reconstruction paradigm.

\begin{figure}
    \centering
    \includegraphics[width=\linewidth]{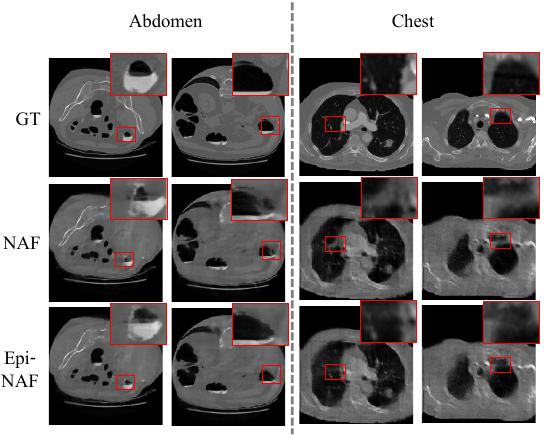}
    \caption{Qualitative comparison between our method (bottom row) and the vanilla NAF (middle row) in the 90\degree\ angle setting, alongside ground-truth slices of abdomen and chest CT scans (top row). We recommend the reader to zoom in electronically for a clearer view of the details.}
    \label{fig:qualitative}
\end{figure}

A qualitative comparison is presented in Figure \ref{fig:qualitative}. Vanilla NAF tends to "hallucinate" high-density regions in areas that should have low density. Our proposed $\mathcal{L}_{\text{ECC}}$ significantly mitigates this issue. This improvement could be crucial in a clinical setting, where the presence or absence of high-density regions in the body can lead to different diagnoses and treatments. 
These qualitative results also reveal the limitations of standard reconstruction metrics in medical imaging: while the PSNR/SSIM improvement of \shortname\ over NAF in the 90\degree\ Abdomen experiment appears modest, the visual improvement in certain regions, as can be seen in the left part of Figure \ref{fig:qualitative}, is substantial and clinically meaningful.

\section{Conclusion}
\label{sec:conclusion}

Neural fields have emerged as a promising paradigm in CT reconstruction, with several methods introduced in recent years that outperform traditional approaches in several challenging settings. In this work, we addressed a critical limitation of neural field-based CT reconstruction, which often produces suboptimal results in limited-angle scenarios. We introduced a novel loss term, $\mathcal{L}_{\text{ECC}}$, grounded in epipolar consistency constraints in X-ray imaging. \shortname, our proposed method, propagates supervision from the available input projections, captured over a restricted angular range, to predicted projections spanning the full angle range required in CBCT setting, effectively regularizing regions that would otherwise remain insufficiently constrained.

Our approach enhances reconstruction performance both quantitatively and qualitatively across various CT scans and limited-angle configurations, providing potentially important benefits for a clinical setting. 

\noindent\textbf{Limitations and Future Work.} Despite our advancements, limited-angle CT remains a significantly under-constrained inverse problem. While our results surpass baseline methods, they still exhibit blurring and loss of important details. In future work, we plan to integrate data-driven priors into the reconstruction framework to further improve reconstruction quality. 
Additionally, we observed that with more fine-tuned hyperparameters, such as the regularization weight 
$\lambda$ and the number of warm-up epochs, we could achieve better results. 
Lastly, $\mathcal{L}_{\text{ECC}}$ is not specific to NAF; it can serve as a plug-and-play regularization term in any neural field-based CT framework, a property we intend to explore in follow-up works.

\section{Compliance with ethical standards}
\label{sec:ethics}

This research study was conducted retrospectively using human subject data made available in open access. Ethical approval was not required as confirmed by the license attached
with the open access data.

\section{Acknowledgments}
Or Litany is a Taub fellow and is supported by the Azrieli Foundation Early Career Faculty
Fellowship. Tzofi Klinghoffer is supported by the National Defense Science \& Engineering Graduate (NDSEG) Fellowship. We thank Dr. Moti Freiman for his important comments.

\bibliographystyle{IEEEbib}
\bibliography{references}

\begin{thebibliography}{10}

\bibitem{quinto2008local}
Quinto et~al.,
\newblock ``Local tomography in electron microscopy,''
\newblock {\em SIAM Journal on Applied Mathematics}, vol. 68, no. 5, pp. 1282--1303, 2008.

\bibitem{zhang2006comparative}
Zhang et~al.,
\newblock ``A comparative study of limited-angle cone-beam reconstruction methods for breast tomosynthesis,''
\newblock {\em Medical physics}, vol. 33, no. 10, pp. 3781--3795, 2006.

\bibitem{schafer2012limited}
Sch{\"a}fer et~al.,
\newblock ``Limited angle c-arm tomography and segmentation for guidance of atrial fibrillation ablation procedures,''
\newblock in {\em MICCAI 2012, Nice, France, October 1-5, 2012, Proceedings, Part I 15}. Springer, 2012, pp. 634--641.

\bibitem{krimmel2005discrete}
S~Krimmel et~al.,
\newblock ``Discrete tomography for reconstruction from limited view angles in non-destructive testing,''
\newblock {\em Electronic Notes in Discrete Mathematics}, vol. 20, pp. 455--474, 2005.

\bibitem{feldkamp1984practical}
Lee~A Feldkamp et~al.,
\newblock ``Practical cone-beam algorithm,''
\newblock {\em Josa a}, vol. 1, no. 6, pp. 612--619, 1984.

\bibitem{gordon1970algebraic}
Gordon et~al.,
\newblock ``Algebraic reconstruction techniques (art) for three-dimensional electron microscopy and x-ray photography,''
\newblock {\em Journal of theoretical Biology}, vol. 29, no. 3, pp. 471--481, 1970.

\bibitem{andersen1984simultaneous}
Andersen et~al.,
\newblock ``Simultaneous algebraic reconstruction technique (sart): a superior implementation of the art algorithm,''
\newblock {\em Ultrasonic imaging}, vol. 6, no. 1, pp. 81--94, 1984.

\bibitem{sidky2006accurate}
Sidky et~al.,
\newblock ``Accurate image reconstruction from few-views and limited-angle data in divergent-beam ct,''
\newblock {\em Journal of X-ray Science and Technology}, vol. 14, no. 2, pp. 119--139, 2006.

\bibitem{niu2014sparse}
Niu et~al.,
\newblock ``Sparse-view x-ray ct reconstruction via total generalized variation regularization,''
\newblock {\em Physics in Medicine \& Biology}, vol. 59, no. 12, pp. 2997, 2014.

\bibitem{huang2013iterative}
Huang et~al.,
\newblock ``Iterative image reconstruction for sparse-view ct using normal-dose image induced total variation prior,''
\newblock {\em PloS one}, vol. 8, no. 11, pp. e79709, 2013.

\bibitem{liu2023dolce}
Liu et~al.,
\newblock ``Dolce: A model-based probabilistic diffusion framework for limited-angle ct reconstruction,''
\newblock in {\em Proceedings of the IEEE/CVF ICCV}, 2023, pp. 10498--10508.

\bibitem{chen2022sam}
Chen et~al.,
\newblock ``Sam’s net: a self-augmented multistage deep-learning network for end-to-end reconstruction of limited angle ct,''
\newblock {\em IEEE Transactions on Medical Imaging}, vol. 41, no. 10, pp. 2912--2924, 2022.

\bibitem{mildenhall2021nerf}
Ben Mildenhall et~al.,
\newblock ``Nerf: Representing scenes as neural radiance fields for view synthesis,''
\newblock {\em Communications of the ACM}, vol. 65, no. 1, pp. 99--106, 2021.

\bibitem{fang2022snaf}
Fang et~al.,
\newblock ``Snaf: Sparse-view cbct reconstruction with neural attenuation fields,''
\newblock {\em arXiv preprint arXiv:2211.17048}, 2022.

\bibitem{cai2024structure}
Cai et~al.,
\newblock ``Structure-aware sparse-view x-ray 3d reconstruction,''
\newblock in {\em Proceedings of CVPR}, 2024, pp. 11174--11183.

\bibitem{zha2022naf}
Zha et~al.,
\newblock ``Naf: neural attenuation fields for sparse-view cbct reconstruction,''
\newblock in {\em MICCAI}. Springer, 2022, pp. 442--452.

\bibitem{zang2021intratomo}
Zang et~al.,
\newblock ``Intratomo: self-supervised learning-based tomography via sinogram synthesis and prediction,''
\newblock in {\em Proceedings of the IEEE/CVF ICCV}, 2021, pp. 1960--1970.

\bibitem{ruckert2022neat}
R{\"u}ckert et~al.,
\newblock ``Neat: Neural adaptive tomography,''
\newblock {\em ACM Transactions on Graphics (TOG)}, vol. 41, no. 4, pp. 1--13, 2022.

\bibitem{aichert2015epipolar}
Aichert et~al.,
\newblock ``Epipolar consistency in transmission imaging,''
\newblock {\em IEEE transactions on medical imaging}, vol. 34, no. 11, pp. 2205--2219, 2015.

\bibitem{grangeat1991mathematical}
Pierre Grangeat,
\newblock ``Mathematical framework of cone beam 3d reconstruction via the first derivative of the radon transform,''
\newblock in {\em Mathematical Methods in Tomography: Proceedings of a Conference held in Oberwolfach, Germany, 5--11 June, 1990}. Springer, 1991, pp. 66--97.

\bibitem{armato2011lung}
Armato III et~al.,
\newblock ``The lung image database consortium (lidc) and image database resource initiative (idri): a completed reference database of lung nodules on ct scans,''
\newblock {\em Medical physics}, vol. 38, no. 2, pp. 915--931, 2011.

\bibitem{biguri2016tigre}
Ander Biguri et~al.,
\newblock ``Tigre: a matlab-gpu toolbox for cbct image reconstruction,''
\newblock {\em Biomedical Physics \& Engineering Express}, vol. 2, no. 5, pp. 055010, 2016.

\bibitem{sidky2008image}
Emil~Y Sidky et~al.,
\newblock ``Image reconstruction in circular cone-beam computed tomography by constrained, total-variation minimization,''
\newblock {\em Physics in Medicine \& Biology}, vol. 53, no. 17, pp. 4777, 2008.

\end{thebibliography}

\end{document}